\documentclass[twocolumn,showpacs,preprintnumbers,amsmath,amssymb,superscriptaddress,prl]{revtex4}
\usepackage{graphicx}% Include figure files
\usepackage{dcolumn}% Align table columns on decimal point
\usepackage{bm}% bold math

\begin{document}

%\preprint{APS/123-QED}

\title{Microwave Spectroscopy of PrOs$_4$Sb$_{12}$: Josephson-Coupled\\ Two-Band Superconductivity and Itinerant $f$ Electrons}% Force line breaks with \\
\author{D.~M.~Broun}
\affiliation{Department of Physics, Simon Fraser University, Burnaby, BC V5A 1S6, Canada}%
\author{P.~J.~Turner}
\author{G.~K.~Mullins}
\author{D.~E.~Sheehy}
\altaffiliation[Present address: ]{Department of Physics, University of Colorado, Boulder, CO 80309, USA}
\affiliation{Department of Physics and Astronomy, University of British Columbia, Vancouver, BC V6T 1Z1, Canada}%
\author{X.~G.~Zheng}
\altaffiliation[Also at: ]{Department~of Physics, Saga University, Saga 840-8502, Japan}
\author{S.~K.~Kim}
\author{N.~A.~Frederick}
\author{M.~B.~Maple}
\affiliation{Department of Physics and Institute for Pure and Applied Physical Sciences, University of California, San Diego, La Jolla, California 92093, USA}%
\author{W.~N.~Hardy}
\author{D.~A.~Bonn}
\affiliation{Department of Physics and Astronomy, University of British Columbia, Vancouver, BC V6T 1Z1, Canada}%

\date{\today}% It is always \today, today,

\begin{abstract}
We report microwave surface impedance measurements made on a single crystal of PrOs$_4$Sb$_{12}$ across the frequency range 0.5 to 21~GHz.  The penetration depth data provide clear thermodynamic evidence for superconductivity arising in two bands coupled by Josephson pair-tunneling, indicating two order parameters with the same symmetry. Detailed conductivity spectra obtained using high-resolution bolometry confirm this picture and extend it, revealing the itinerant nature of the Pr $f$ electrons through the observation of quasiparticles with heavily renormalized mass.
\end{abstract}

\maketitle

Interest in the recently discovered skutterudite superconductor PrOs$_4$Sb$_{12}$ has focused on the possibility that this compound could provide the first example of superconducting electron pairing mediated by the exchange of quadrupolar charge density fluctuations  \cite{bauer02,maple02}.  Based on magnetic susceptibility, specific heat, and neutron diffraction measurements, it has been suggested that the Pr$^{3+}$ ground state in the cubic crystal field is a nonmagnetic doublet that carries an electric quadrupole moment, separated by $\sim 7$~K from a low lying triplet \cite{bauer02,maple02}.  This raises the interesting possibility of an electric quadrupolar Kondo effect \cite{cox87}, and specific heat $C(T)$ measurements indicate the formation of a heavy fermion metallic state \cite{bauer02,maple02}.  However, it should be noted that several investigators have concluded that the ground state of Pr$^{3+}$ is a singlet \cite{aoki02,kohgi03}.  Evidence for unconventional superconductivity in PrOs$_4$Sb$_{12}$ includes a double transition in $C(T)$\cite{maple02,vollmer03}, a report of multiple superconducting phases of different symmetry in field-dependent thermal conductivity $\kappa(T,H)$ \cite{izawa03}, the observation of low temperature power laws in $C(T)$ \cite{bauer02,maple02} and London penetration depth $\lambda_{\textrm L}(T)$ \cite{chia03}, and spontaneous magnetic fields appearing below the superconducting transition temperature $T_{\textrm c}$, indicating time reversal symmetry breaking \cite{aoki03}.  The occurrence of a high field ordered phase (HFOP) between 4.5~T and 16~T below 1~K has been inferred from electrical resistivity \cite{maple02,ho02,maple03}, specific heat \cite{aoki02,vollmer03}, thermal expansion \cite{oeschler03}, and magnetization \cite{ho02,tenya03,tayama03} measurements.  From neutron scattering studies in high magnetic fields it was concluded that antiferro-quadrupolar order occurs in the HFOP \cite{kohgi03}.  
In this Letter we present the first measurements of the microwave surface impedance $Z_{\textrm s}$ of PrOs$_4$Sb$_{12}$, from which we infer a rapid drop in quasiparticle scattering below $T_{\textrm c}$, the existence of Josephson-coupled two-band superconductivity (implying two order parameters of the {\em same} symmetry), and heavy, itinerant quasiparticles of mass $\sim 100~m_{\textrm e}$.

High quality single crystals of PrOs$_4$Sb$_{12}$ were grown from an Sb flux \cite{bauer01}.  Powder x-ray diffraction indicates that the samples are single phase, with a stoichiometric ratio of 1 Pr atom per formula unit \cite{ho02}.  Sharp superconducting transitions in resistivity, susceptibility and microwave conductivity indicate high homogeneity, and the low residual resistivity ($< 5~\mu\Omega$cm) attests to long carrier mean free paths and low levels of disorder \cite{bauer02}.  The sample used in our experiment was an as-grown single crystal measuring 0.18~mm$\times$0.22~mm$\times$1.1~mm, with clean, shiny faces aligned to the cubic crystallographic axes.   Measurements of $Z_{\textrm s}$ were made using three sets of apparatus cooled in $^4$He cryostats.  A loop--gap resonator (954~MHz) \cite{hardy93} and a dielectric resonator (5.49~GHz) \cite{morgan03} were used to make cavity perturbation measurements of the surface resistance $R_{\textrm s}$ and changes in surface reactance $\Delta X_{\textrm s}(T)$ from 1.16~K to 120~K.  Frequency scans of  $R_{\textrm s}$ across the range 0.5~GHz to 21~GHz were performed using a bolometric surface resistance probe \cite{turner03b}, at a set of fixed temperatures.  The $Z_{\textrm s}$ data have been analyzed to obtain the microwave conductivity $\sigma = \sigma_1 - {\textrm i}\sigma_2$ by assuming the local electrodynamic relation $Z_{\textrm s} \equiv R_{\textrm s} + {\textrm i}X_{\textrm s} = \sqrt{{\textrm i}\omega\mu_0/\sigma}$.  A long penetration depth $\lambda_{\textrm L} \approx 3000$~\AA\ (this work and Ref.~\onlinecite{maclaughlin02}) and a short coherence length $\xi_0 \approx 120$~\AA \cite{bauer02} make local electrodynamics a valid assumption. 

\begin{figure}
\includegraphics[width=85mm]{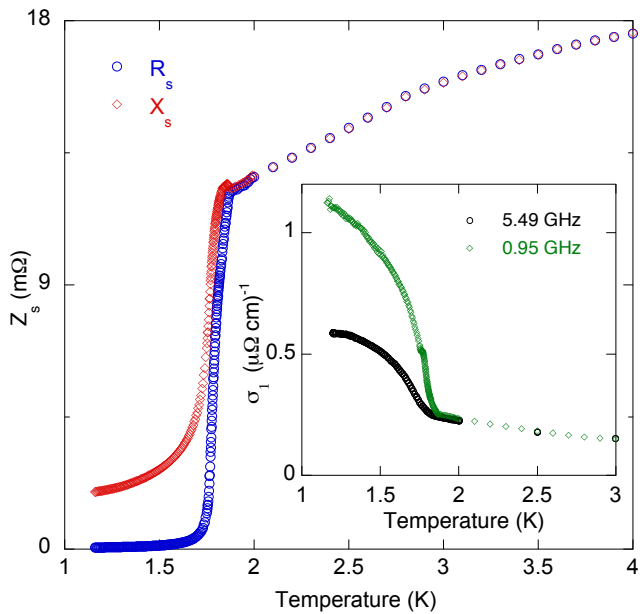}% Here is how to import EPS art
\caption{\label{fig1}  (color online). Microwave surface resistance $R_{\textrm s}$ and reactance $X_{\textrm s}$ at 954~MHz.  Inset: Quasiparticle conductivity $\sigma_1(T)$ data  at 954~MHz and 5.49~GHz indicate a rapid collapse in quasiparticle scattering below $T_{\textrm c}$.}
\end{figure}

Fig.~\ref{fig1} shows the $R_{\textrm s}$ and $X_{\textrm s}$ data at 954~MHz.  Above $T_{\textrm c} = 1.872$~K, $R_{\textrm s}(T)$ and $\Delta X_{\textrm s}(T)$ have the same temperature dependence, as expected in the classical skin-effect regime where $R_{\textrm s} = X_{\textrm s}$.  Enforcing this condition by adding a temperature independent offset to $\Delta X_{\textrm s}(T)$ allows us to determine the {\em absolute} reactance, which here corresponds to $\lambda_{\textrm L}(T=1.2$~K$) = 2650$~\AA.  The strong temperature dependence of $R_{\textrm s}$ and $X_{\textrm s}$ in the normal state indicates the presence of  substantial inelastic scattering. 
Below $T_{\textrm c}$ there is a sharp drop in $R_{\textrm s}$ followed by a gradual decrease to a residual value of 50~$\mu\Omega$.  Similar features are observed at 5.49~GHz, and from $Z_{\textrm s}$ we obtain $\sigma_1(T)$ at the two frequencies (see inset of Fig.~\ref{fig1}).  The frequency dependence of $\sigma_1$ is weak in the normal state, consistent with a quasiparticle scattering rate much higher than microwave frequencies.  Below $T_{\textrm c}$ the rapid rise in $\sigma_1(T)$ and the development of strong frequency dependence together point to a sudden drop in scattering.  This is a characteristic feature of strongly correlated metals such as the high $T_{\textrm c}$ cuprates \cite{bonn92}, in which the opening of the superconducting gap quickly suppresses the electronic fluctuations responsible for strong normal state inelastic scattering.

\begin{figure}
\includegraphics[width=85mm]{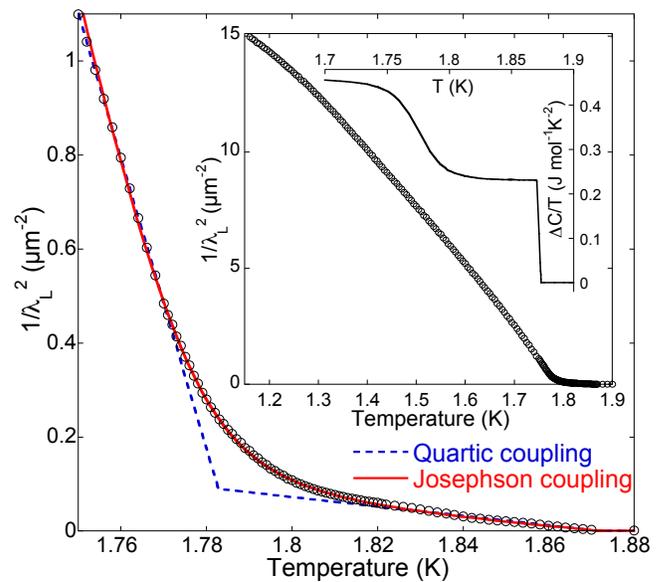}% Here is how to import EPS art
\caption{\label{fig2} (color online).  London penetration depth $\lambda_{\textrm L}$ plotted as $1/\lambda_{\textrm L}^2(T)$.  Fits to the data using a two band G--L model with Josephson coupling (solid line) and quartic coupling (dashed line) are shown.  Inset 1: $1/\lambda_{\textrm L}^2(T)$ over a wider temperature range. Inset 2: Superconducting contribution to $C/T$ calculated from the Josephson-coupled free energy. }
\end{figure}

\begin{figure}
\includegraphics[width=85mm]{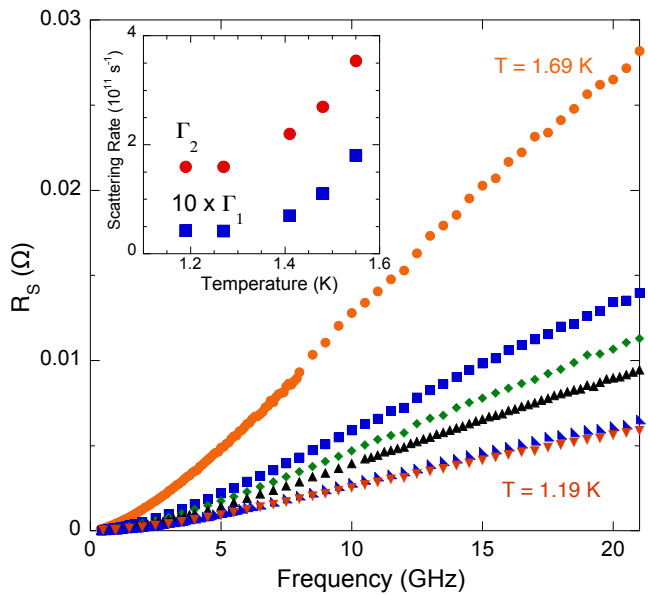}% Here is how to import EPS art
\caption{\label{fig3}  (color online). Surface resistance spectra $R_{\textrm s}(\omega)$ taken at different temperatures.  From top to bottom, T = 1.69~K, 1.55~K, 1.48~K, 1.41~K, 1.27~K and 1.19~K. Inset: scattering rates $\Gamma_1$ (multiplied by 10) and $\Gamma_2$ inferred from fits to $\sigma_1(\omega)$. }
\end{figure}

Fig.~\ref{fig2} shows the 954~MHz $\sigma_2(T)$ data, plotted as $1/\lambda_{\textrm L}^2(T) \equiv \omega \mu_0 \sigma_2(T)$.   At $T_{\textrm c} = 1.872$~K, $1/\lambda_{\textrm L}^2(T)$ has a sharp onset, less than 5~mK wide, below which it increases linearly with a slope of $-1.1~\mu$m$^{-2}$K$^{-1}$.  A rounded change in slope occurs 70~mK below $T_{\textrm c}$, followed by a second regime of linear $T$ dependence with a slope of $-31~\mu$m$^{-2}$K$^{-1}$.  This behaviour is suggestive of two superconducting transitions and is consistent with the observation of double transitions in $C(T)$ \cite{maple02,vollmer03} and  $\kappa(T,H)$ \cite{izawa03}.   To model $\lambda_{\textrm L}$, a thermodynamic quantity that can be derived from the free energy, we consider the case of two weakly coupled bands. The Ginzburg--Landau (G--L) free energy density $f$ can be written $f = \sum_{ i = 1}^2 f_i + f_{\textrm m} + f_{\textrm c}$, where the intraband contributions $f_i = 1/(2 m^\ast_i) \left|(\textrm{i} \hbar \bm{\nabla} + 2 e \bm{A})\psi_i\right|^2 + \alpha_i(T) \left|\psi_i\right|^2 + \beta_i/2 \left|\psi_i\right|^4$ with  $\alpha_i(T) = \alpha_{0i} (T - T_{{\textrm c}i})$ and $\beta_i =$~const.  The magnetic term is $f_m = (\bm{\nabla} \times \bm{A})^2/(2 \mu_0)$, and the interband couplings allowed by symmetry can be written $f_c = - \eta_1 \left|\psi_1\right| \left|\psi_2\right| \cos\phi + \eta_2\left|\psi_1\right|^2\left|\psi_2\right|^2 (1 + a \cos 2\phi)$, with $\phi$ the phase difference between $\psi_1$ and $\psi_2$.  Note that $\eta_1$ is the strength of the interband Josephson tunneling and can be nonzero only if $\psi_1$ and $\psi_2$ have the {\em same} symmetry.  The quartic coupling is insensitive to order parameter symmetry and contains a $\cos 2 \phi$ term originating from {\em double} pair transfer.  (Depending on the sign and magnitude of $a$, there can be a transition from phase aligned ($\phi = 0$) to phase rotated ($\phi = \frac{\pi}{2}$) states, but the data do not support this and $\phi = 0$ is therefore assumed in the following.) Minimizing the free energy with respect to magnetic field $\bm{B} = \bm{\nabla} \times \bm{A}$ gives the London equation for the supercurrent density, from which we obtain the two-band expression for the penetration depth: $\lambda_{\textrm L}^{-2} = 4 \mu_0 e^2 \{\frac{\left|\psi_1\right|^2}{m^\ast_1} + \frac{\left|\psi_2\right|^2}{m^\ast_2}\}$, where  $\left|\psi_i\right|^2$ are the pair densities and $m_i^\ast$ the pair effective masses.  Therefore, since the build-up of superfluid density below the second transition in PrOs$_4$Sb$_{12}$ occurs about 30 times faster than the initial increase, we have either a small pair density in the first band or a large pair mass.  In the case $\eta_1 = \eta_2 = 0$, the model reduces to that of two independent superconductors with transition temperatures $T_{{\textrm c}1}$ and $T_{{\textrm c}2}$.  In that case the usual results follow on minimizing $f$ with respect to $\psi_i^\ast$: $\left|\psi_i\right|^2 = \alpha_{0i} (T_{{\textrm c}i} - T)/\beta_i$ and $\Delta C_i/T = - \left(\partial^2f/\partial T^2\right)_{T=T_{{\textrm c}i}} = \alpha_{0i}^2/\beta_i$.  That is, in each band the pair density builds up linearly with decreasing temperature, with a double jump in the specific heat $C$.  However, only the $\eta_1$ coupling can give the delicate curvature seen in Fig.~\ref{fig2}.  
 To show that we consider the cases of Josephson and quartic couplings separately and, by solving the G--L equations numerically, fit to $1/\lambda_{\textrm L}^2(T)$.  Several parameters of the G--L model are fixed by other experiments: the two heat capacity jumps in a mosaic of crystals from the same batch were measured to be $\Delta C_1/T \approx \Delta C_2/T = 0.23$~J.mol$^{-1}$K$^{-2}$ \cite{maple02}, which fix $\alpha_{0i}^2/\beta_i$; the effective mass of the pairs in the second band is taken to be twice the average cyclotron mass inferred from dHvA experiments, $m_2^\ast \approx 9~m_{\textrm e}$ \cite{sugawara02}; and $T_{{\textrm c}1}$, which is unaffected by the strength of the interband coupling, is set to 1.872~K. Good initial guesses for the ratios $\alpha_{0i}/\beta_i$ can be obtained from the slope of  $1/\lambda_{\textrm L}^2(T)$ in the linear regions and, anticipating the results of the conductivity analysis below, we set $m_1^\ast = 30~m_2^\ast$.  (Heavy-mass quasiparticles have not been detected in the dHvA experiments \cite{sugawara02}, but they may be localized in the HFOP above 4.5~T \cite{aoki02,vollmer03,kohgi03}.  Analyses of specific heat and upper critical field data provide indirect evidence for an average quasiparticle mass of order 50~$m_{\textrm e}$ \cite{bauer02}.   In any case, the choice of mass ratio $m_1^\ast/m_2^\ast$ does not affect the quality of the fits to the $\lambda_{\textrm L}$ data, nor the conclusions drawn about the nature of the interband coupling.)  The best fit to the data in the case of pure Josephson coupling ($\eta_2 = 0$) is shown by the solid curve in Fig.~\ref{fig2} and corresponds to setting $T_{{\textrm c}2} = 1.763$~K, $\alpha_{01}/k_{\textrm B} = 0.024$, $\alpha_{02}/k_{\textrm B} = 0.012$, and $\eta_1 = 0.0053 \sqrt{\alpha_{01}\alpha_{02}} T_{{\textrm c}1}$, where the band-dependent G--L parameters $\kappa_i = \sqrt{m_i^{\ast 2} \beta_i/2 \mu_0 e^2 \hbar^2}$ are $\kappa_1 = 99$ and $\kappa_2 = 1.72$. The pair-tunneling solution matches $1/\lambda_{\textrm L}^2(T)$ extremely well, properly capturing the gentle rounding that occurs in the vicinity of $T_{{\textrm c}2}$. We can interpret this as an avoided crossing of $\psi_1$ and $\psi_2$ caused by mixing of states of the {\em same} symmetry.  Shown by the dashed curve in Fig.~\ref{fig2} is the case of pure quartic coupling ($\eta_1 = 0, \eta_2 (1 + a) = 0.22 \sqrt{\beta_1\beta_2}$), appropriate for order parameters of {\em different} symmetry.  In that case we see level crossing, not level repulsion, for {\em any} value of $\eta_2$. This is strong evidence that $\psi_1$ and $\psi_2$ have the same symmetry, and that the second phase transition reported in other experiments \cite{izawa03,vollmer03} is in fact a cross over. ($\psi_2$ is finite at all temperatures below $T_{{\textrm c}1}$.) We believe the apparent discrepancy with the $\kappa(T,H)$ result --- that the nodal structure of the order parameter changes on crossing $T_{{\textrm c}2}$ --- could be resolved by taking into account  the sudden drop in quasiparticle scattering on cooling through $T_{{\textrm c}1}$, and the rapid build-up of $1/\lambda_{\textrm L}^2(T)$ below $T_{{\textrm c}2}$.  (The average quasiparticle Doppler shift from vortex flow fields is proportional to $1/\lambda_{\textrm L}^2(T)$ \cite{yu95}.)

\begin{figure}
\includegraphics[width=85mm]{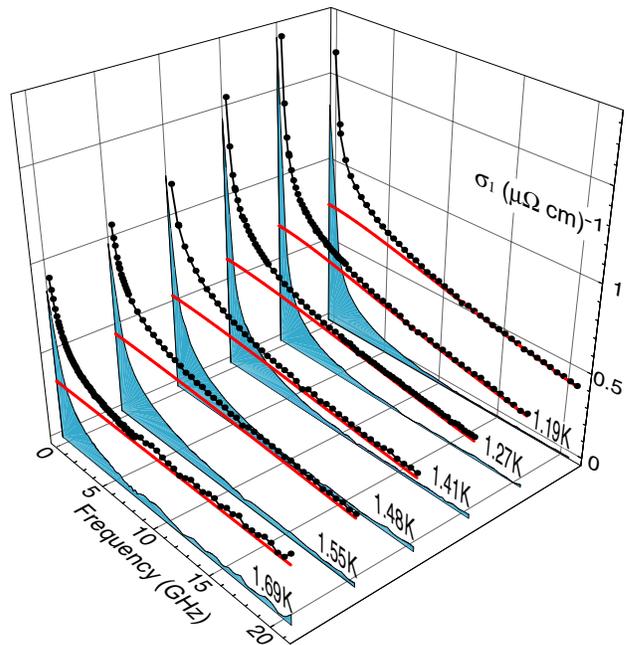}% Here is how to import EPS art
\caption{\label{fig4}  (color online).  Solid points show the $\sigma_1$ data obtained from a Kramers--Kr\"onig analysis of $R_{\textrm s}(\omega,T)$ and $1/\lambda_{\textrm L}^2(T)$.  Solid lines show the modified Drude fits to the broad background.  The shaded regions underneath are the difference between the $\sigma_1$ data and the broad background fits.  }
\end{figure}

An unconventional two-band superconductor should exhibit novel physical properties.  Leggett has predicted a new collective mode, corresponding to Josephson plasma resonance between the bands \cite{leggett66}.  This mode is expected to be difficult to observe experimentally, as it does not perturb the long wavelength charge density and hence couples weakly to electromagnetic radiation.  (Our attempts to excite the mode with microwaves have been unsuccessful, with no indication of collective mode absorption in $R_s(\omega,T)$.)  Babaev  has studied the magnetic properties of two-band superconductors and concludes that these systems can contain fractional vortices arising from topological defects in only one of the order parameters \cite{babaev02}.  In addition, Gurevich and Vinokur predict static and dynamic solotonic phase textures that should appear in nonequilibrium transport experiments \cite{gurevich03}.

$R_{\textrm s}(\omega,T)$ is plotted from 0.5~GHz to 21~GHz in Fig.~\ref{fig3}, at six different temperatures in the superconducting state.  With no sign of resonant absorption by collective excitations, we expect $R_{\textrm s}$ to be due solely to thermally excited quasiparticles.  To extract the quasiparticle conductivity spectra $\sigma_1(\omega,T)$ from $R_s(\omega,T)$ and $1/\lambda_{\textrm L}^2(T)$ we apply a Kramers--Kr\"onig analysis \cite{turner03b,turner03a}, which properly accounts for quasiparticle contributions to the screening of electromagnetic fields.  The conductivity spectra are plotted in Fig.~\ref{fig4}, for temperatures between 1.19~K and 1.69~K.  At each temperature, $\sigma_1(\omega)$ has a sharp, cusp-like upturn at low frequencies with a broad high frequency tail.  With decreasing $T$ the spectra sharpen and increase in height as they draw in more spectral weight from high frequencies.  Broadly similar behaviour has been observed in high quality samples of Ortho-II ordered YBa$_2$Cu$_3$O$_{6.5}$, where it was concluded that the cusp-like lineshapes and strongly temperature-dependent linewidths were close to the behaviour expected for weak (Born-limit) scattering in a superconductor with energy gap nodes and an energy dependent scattering rate \cite{turner03a}.   In that work a modified Drude  spectrum $\sigma_1(\omega) = \sigma_0/(1 + (\omega/\Gamma)^y)$, with $y \approx 1.45$, was shown to capture the phenomenology of energy dependent scattering.    We have attempted to fit this form to our  PrOs$_4$Sb$_{12}$ data, with no success.  The problem lies in the combination of a very sharp low frequency peak and a broad high frequency background.  The situation improves, however, if we allow for the possibility of two bands with different scattering rates and fit the sum of two modified Drude spectra at each temperature. In Fig.~\ref{fig4} the broad spectra are plotted as solid lines, which have then been subtracted from the data points to reveal the narrow spectra, represented by the shaded regions.   In all the fits the value of $y$ is close to 1.50 for both the broad and narrow components.  The values of $\sigma_0$ are the zero frequency intercepts, and the values of scattering rates are plotted in the inset of Fig.~\ref{fig3} as $\Gamma_1$ and $\Gamma_2$.  Both $\Gamma_1$ and $\Gamma_2$ decrease strongly on cooling, as inferred from the resonator data in  Fig.~\ref{fig1}. $\Gamma_1$ is about a factor of 30 smaller than $\Gamma_2$ over the whole temperature range: it is this fact that allows a unique separation of the spectra.  The $T$ dependent spectral weight $\int_0^\infty \sigma_1(\omega,T) {\textrm d}\omega$ is in good accord with the G--L model for $1/\lambda_{\textrm L}^2(T)$ if the narrow spectra are identified with band 1 and the broad spectra with band 2.  To understand the form of the unusual $\sigma_1(\omega,T)$ spectra better we look to Fermi-liquid transport theory where, for the case of electron--phonon coupling, Prange and Kadanoff have found that interactions do not renormalize the low-frequency transport coefficients, due to a cancellation  between the renormalizations of the lifetime $\tau^\ast$ and mass $m^\ast$ \cite{prange64}.  That is, as  $m^\ast$ increases, the width of the conductivity spectrum shrinks, but the dc conductivity remains fixed, so that the total conductivity spectral weight  $\propto 1/m^\ast$.  Varma has argued that sufficiently local interactions will have the same effect in heavy fermion metals \cite{varma85}, where $m^\ast \sim 100~m_{\textrm e}$, implying a reduction in quasiparticle scattering rate by a similar factor.  
From this we infer $m_1^\ast/m_2^\ast \approx \Gamma_2/\Gamma_1 \approx 30$  in PrOs$_4$Sb$_{12}$, motivating our parameter choice in the G--L model.  The conductivity spectra therefore provide direct evidence for heavy quasiparticles and hence of the itinerant nature of the Pr $f$ electrons outside the high field ordered phase.

In summary, we have used microwave measurements as a thermodynamic and spectroscopic probe of PrOs$_4$Sb$_{12}$, finding evidence for novel two-band superconductivity.  The bands are coupled by Josephson pair tunneling, requiring that there be two order parameters with the {\em same} symmetry.  Quasiparticle conductivity spectra confirm this and provide a unique insight into the transport dynamics, revealing a collapse in scattering below $T_{\textrm c}$, and the presence of heavy itinerant quasiparticles.

We gratefully acknowledge useful discussions with I.~Affleck, M.~Franz, I.~F.~Herbut, B.~Morgan and D.~C.~Peets, and financial support from the National Science and Engineering Research Council of Canada, the Canadian Institute for Advanced Research, and the U.~S. Department of Energy and National Science Foundation.

\end{document}